\documentclass[preprint,prl,showpacs,byrevtex,superscriptaddress]{revtex4-1}
\usepackage{graphicx}
\usepackage{epsfig}

\usepackage[caption=false]{subfig}
\usepackage{amsmath,braket,relsize}
\usepackage{amssymb}
\usepackage{bm}
\usepackage{color}

\begin{document}

\title{Theory of Graphene-based Plasmonic Switch}
\author{Kyungsun Moon$^*$}
\author{Suk-Young Park}
\affiliation{Department of Physics and IPAP, Yonsei University, Seoul 03722, Korea, *E-mail: kmoon@yonsei.ac.kr}

\begin{abstract}

We have theoretically studied a graphene-based plasmonic waveguide, which can gate the transmission
of a surface plasmon polariton (SPP) localized near the graphene-semiconductor interface.
When a gate voltage is applied above a certain critical value, the charge density modulation in the quasi two-dimensional electron gas
formed in the inversion layer can induce a local plasma resonance.
Since the local plasma resonance is strongly coupled to the SPP, it can suppress the transmission of the SPP.
By calculating the propagation length of the SPP with varying gate voltage, we have obtained
the sharp switching line shape. We have demonstrated that the wavelength of the SPP can be reduced below $\sim1/100$ of that of an incident light
and the propagation length of the SPP can be significantly reduced by a factor of $\sim15$ upon switching.
This ensures that our plasmonic waveguide can operate effectively as a plasmonic switch for the SPP.
\end{abstract}

\maketitle

Recent technological advances in dynamically controlling the plasmon properties of photoactive
materials have played an important role from data processing and transmission to energy harvesting\cite{krasavin,nikolajsen,ragip,matthew,jennifer,abb}.
Surface plasmon polaritons are electromagnetic excitations propagating at the interface between a dielectric and a metal,
evanescently confined in the perpendicular direction. These electromagnetic surface waves arise due to
the coupling of the electromagnetic fields in the dielectric to the charge density fluctuations on the metal surface.
Since their wavelengths are typically much shorter than that of an incident light, they have been successfully applied to
the sub-wavelength optics in microscopy and lithography beyond the diffraction limit\cite{zeng,barnes,huidobro,ozbay,akimov}.
Photonic devices have great advantages over electronic ones due to the fast operation speed. However, large device size has remained
as an obstacle for practical purpose. Hence the SPP can provide a breakthrough for miniaturization of photonic devices.
Extensive studies have been performed to investigate the SPP dispersion relations for various waveguide structures both theoretically and experimentally\cite{jingchen,jesly,tjdavis,xinil}.
For practical photonic device applications, high Q-factor of the SPP is required and hence much efforts have been taken to increase the Q-factor such as
optimizing waveguide geometry and adopting low-loss materials\cite{myungki}.
In order to control the SPP in a sub-wavelength photonic device, various methods have been proposed such as optical controls using heat, gate voltage, and light\cite{krasavin,nikolajsen,ragip,matthew,jennifer,abb}.
Since the coupling strength between the SPP and external fields such as gate voltage is relatively weak, they usually require high switching voltage and
long channel length. Moreover, slow switching time typically shown in previous experiments is a great challenge to overcome in order to achieve
high-speed information processing.

Graphene, a two-dimensional sheet of carbon atoms placed on the honeycomb lattice, has attracted a lot of theoretical and experimental attentions
due to its superior physical properties such as high mobility, elastic, thermal, and optical properties\cite{akgeim,ksnovoselov,akgeim2,changgulee,alexanderabalandin}.
The optical absorption rate of graphene has exhibited a universal value of $\pi\alpha\cong2.3\%$ in the visible and near-infrared region, where $\alpha$ denotes the fine structure constant. This broadband absorption characteristic has been utilized for the graphene-based photodetectors/modulators\cite{Constant,Wong}.
In addition, the graphene plasmonic effects have been investigated for different applications ranging from light modulation to biological/chemical sensing\cite{
low,jablan,zfei,diaz1,grigorenko,koppens,ferreira,diaz2}. The two-dimensional nature and high mobility of graphene have enabled an enhanced light-matter interactions and fast photo-detection\cite{Echtermeyer,Jadidi,Fernandez}.

\begin{figure}
\epsfxsize=6in \epsfysize=5in \epsffile{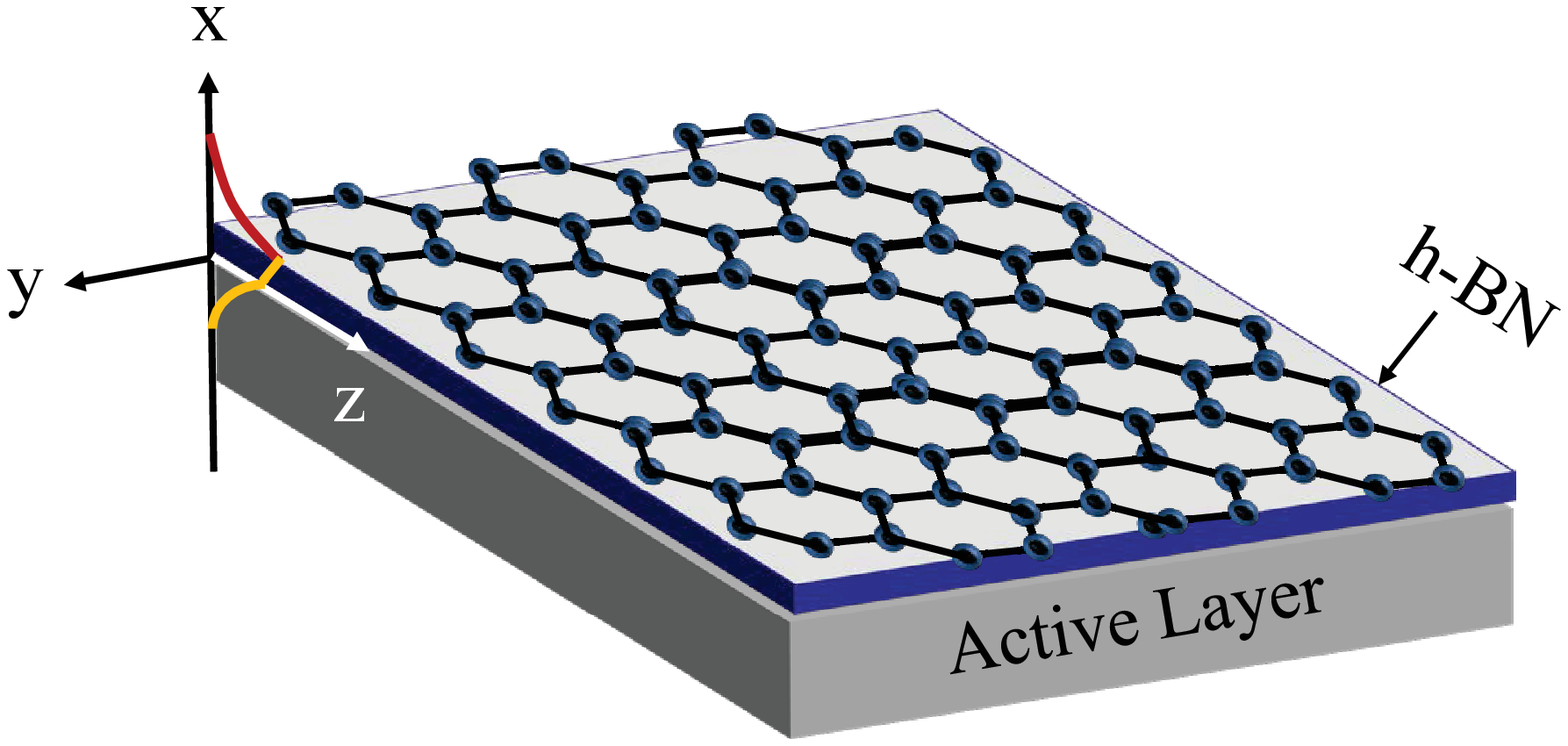}
\caption{
The schematic of our plasmonic waveguide. Hexagonal boron nitride film is used as a gate dielectric material for graphene.
The active layer can be taken to be a p-type semiconductor with high mobility for fast switching such as a p-type GaAs.
The SPP propagates along the $z$-axis.}
\label{Figure1}
\end{figure}

\begin{figure}
\epsfxsize=6in \epsfysize=3.5in \epsffile{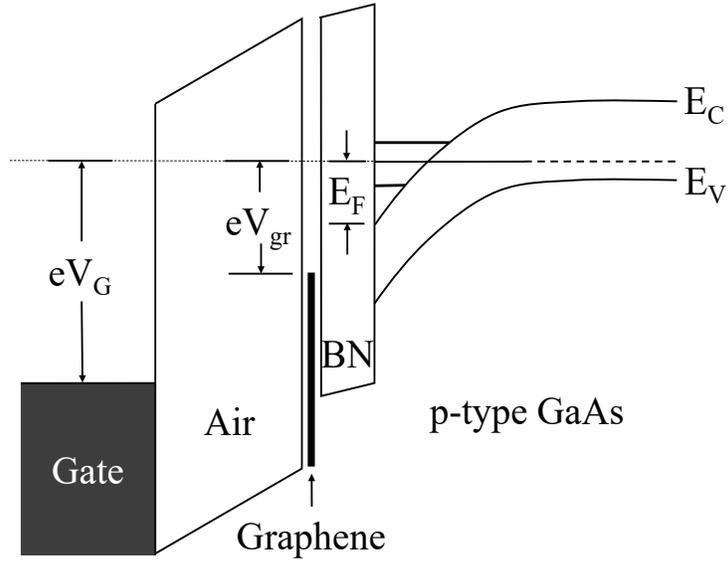}
\caption{
The energy band diagram of the plasmonic waveguide. Graphene is separated from the top gate by an air gap. The voltage $V_{\rm gr}$ at graphene and the top gate voltage $V_{\rm G}$ can control independently the Fermi energy of graphene and the electrostatic potential energy profile for the band bending.}
\label{Figure1_1}
\end{figure}

In the paper, we have studied the plasmonic waveguide, which controls the transmission of an incident light by switching the coupling of the SPP localized at the graphene-semiconductor interface. The schematic of our device is illustrated in Fig. \ref{Figure1}. Hexagonal boron nitride (h-BN) film is used as a gate dielectric material for graphene.
Graphene-based electronic devices have been actively developed in order to utilize an excellent electrical conduction characteristics of graphene. However, most of the experimental results do not show dramatic increase in the device efficiency as compared to the theoretical prediction.
As a van der Waals material, h-BN has performed well as a perfect substrate material for graphene, which can protect the high mobility of graphene similar to the level of the suspended graphene flake\cite{dean}. The active layer will be taken to be a p-type semiconductor with high mobility in order to achieve a fast switching such as a p-type GaAs.
The energy band diagram of the plasmonic waveguide is shown in Fig. \ref{Figure1_1}. The top gate is separated from graphene by an air gap\cite{Gorbachev}. By varying the voltage $V_{\rm gr}$ at graphene and the top gate voltage $V_{\rm G}$, one can control independently the Fermi energy of graphene and the electrostatic potential energy profile $V(z)$ for band bending.

When the positive top gate voltage is applied above a certain threshold value, an inversion layer will be formed within the depletion region of a p-type semiconductor and quasi two-dimensional electron gas will appear in the inversion layer. By varying the top gate voltage further, the total number of electrons in the 2D electron gas will increase and the electron number density profile can be spatially modulated as well. Above a certain critical voltage, it can induce a local plasma resonance in the inversion layer.
Due to the strong coupling of the SPP to the local plasma resonance, the transmission of the SPP will be strongly suppressed in our plasmonic waveguide. As a specific example, we have chosen a p-type GaAs as an active layer. For the experimentally available set of parameters, we have calculated that for an incident light with mid-infrared wavelength $\lambda=8\mu m$, the wavelength of the SPP can be reduced below $\sim1/100$ of that of an incident light. Upon approaching the threshold voltage, the propagation length of the SPP has been dramatically reduced by a factor of $\sim10$ upon switching. Hence our plasmonic switch can effectively gate the transmission of the SPP by applying a gate voltage. Since the gating operation will reflect upon the underlying resonance phenomenon, our plasmonic waveguide demonstrates much sharper switching line shape in comparison to the other devices\cite{diaz1}. It has been experimentally demonstrated that a pumped laser pulse can modulate the carrier density by about $6\%$ in the ITO/nano-antennae hybrid structure\cite{abb}. For the future application to all-optical plasmonic switch, our device has sharpness characteristic, which can be operative to such a modest density modulation.

We will investigate the physical properties of the SPP in our plasmonic waveguide.
The magnetic field in the transverse magnetic (TM) mode of our waveguide can be described by ${\vec H}(x)=\tilde{H}(x)\exp\left[i(k_{z}z-\omega t)\right]{\hat y}$, which propagates along the $z$-axis with frequency $\omega$ and wave number $k_z$. It will satisfy the following non-uniform wave equation\cite{ginzburg,akimov}
\begin{equation}
\frac{\partial^2\tilde{H}}{\partial x^2}-\frac{1}{\epsilon(\omega,x)}\frac{d\epsilon(x)}{dx}\frac{\partial\tilde{H}}{\partial x}-\left[k^2_{z}-k^{2}\epsilon(\omega,x)\right]\tilde{H}=0,
\label{22202}
\end{equation}
where $k$ represents the wave number of the incident light in free space and $\epsilon(\omega,x)$ the relative permittivity at frequency $\omega$ as a function of position $x$.
The impedance of a waveguide can be defined by $Z=E_{z}/H_{y}$ with $E_z=\left(i/k\epsilon(\omega,x)\right)\left(\partial H_y/{\partial x}\right)\sqrt{\mu_0/\epsilon_0}$.
The dimensionless impedance $Z(x)$ in units of $\sqrt{\mu_0/\epsilon_0}$ will obey the following non-uniform wave equation, which can be written by
\begin{equation}
\frac{\partial Z}{\partial x}+ik\left[1-\epsilon(\omega,x)Z^2-\frac{k^2_{z}}{k^2\epsilon(\omega,x)}\right]=0.
\label{impedance}
\end{equation}
In our plasmonic waveguide, we have taken a p-type semiconductor as an active layer, whose relative permittivity $\epsilon(\omega,x)$ can be spatially modulated by applying the top gate voltage.
According to the Drude model, $\epsilon(\omega,x)$ in the active layer will depend on the local plasma frequency $\omega_p(x)$ as follows
\begin{equation}
\epsilon(\omega,x)=\epsilon_{\infty}-{\frac{\omega^2_{p}(x)}{\omega(\omega+i\nu)}}
\end{equation}
where $\omega_p(x)$ depends on the local electron density $n(x)$ through $\omega_{p}(x)=\sqrt{n(x)e^2/m_c^{*}\epsilon_0}$ with $m_c^{*}$ being the effective mass for conductivity and $\epsilon_{\infty}, \nu$ represent the high-frequency relative permittivity and the plasma damping rate of the active layer, respectively.

\begin{figure}
\epsfxsize=7in \epsfysize=4in \epsffile{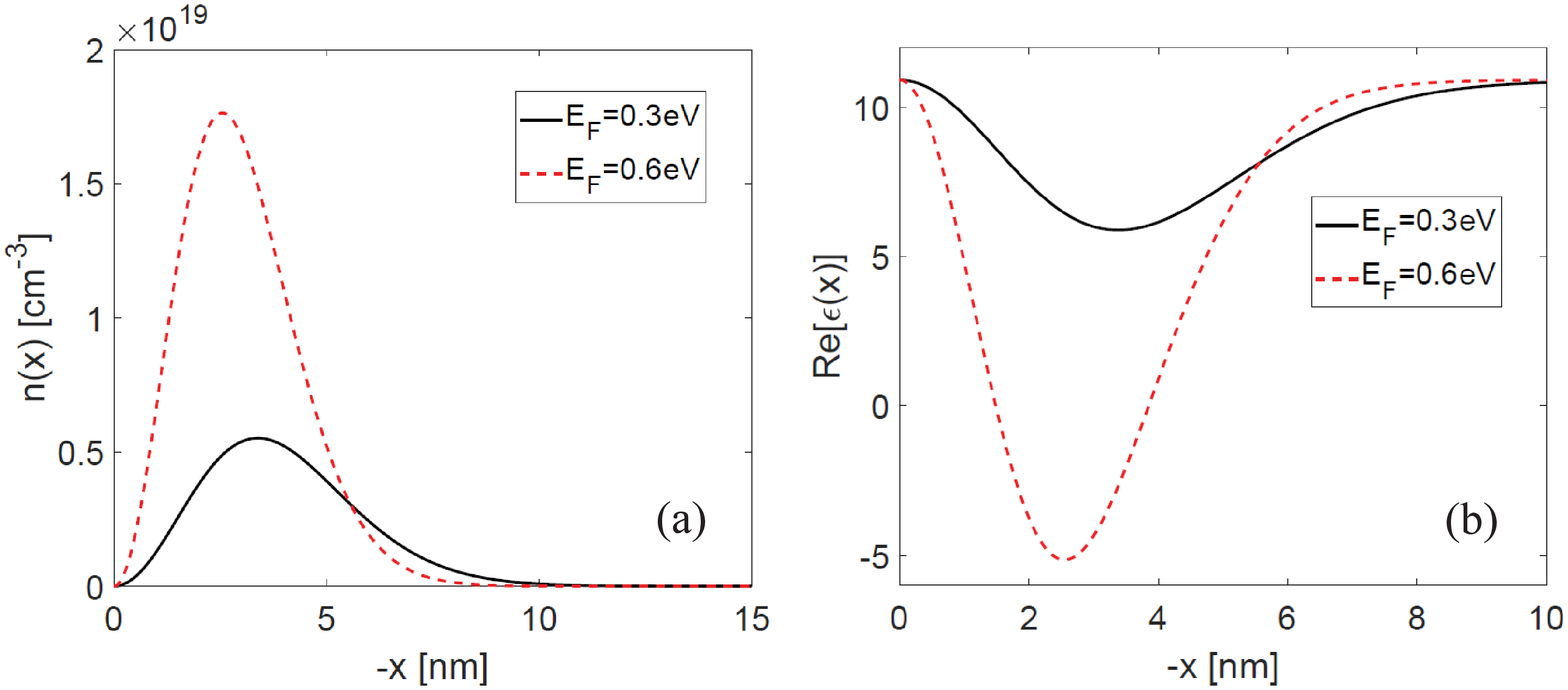}
\caption{
(a) The electron number densities $n(x)$ are plotted as a function of $-x$ for $E_{\rm F}=0.3, 0.6{\rm eV}$. (b) The real parts of dielectric permittivities $Re\left[\epsilon(x)\right]$ are plotted as a function of $-x$ for $E_{\rm F}=0.3, 0.6{\rm eV}$.}
\label{Figure1_2}
\end{figure}

Now we will explain how to obtain the electron number density profile $n(x)$ as a function of the Fermi energy $E_{\rm F}$ of the quasi two-dimensional electron gas. Electrons moving in a quantum well with potential energy $V(x)$ will satisfy the following Schr{\"o}dinger equation
\begin{equation}
-{\hbar^2\over 2 m^*_x}\frac{d^2\Psi_n(x)}{d x^2}+\left(V(x)-E_n\right)\Psi_n(x)=0,
\label{Schroedinger}
\end{equation}
where $m^*_x$ represents the effective mass along the {$x$}-axis. The electrostatic potential $\Phi(x)=-V(x)/e$ will satisfy Gauss's law
\begin{equation}
\frac{d^2\Phi(x)}{d x^2}={e\over \epsilon_s}\left[n(x)-p(x)+N_a\right],
\label{Poisson}
\end{equation}
where $\epsilon_s, n(x), p(x)$ represent the dielectric permittivity of the semiconductor, the electron number density, and the hole number density, respectively, and $N_a$ the doping concentration of acceptor impurities. One can assume that $p(x)$ is negligibly small in the depletion region\cite{MOSFET}.
For the low energy eigenstates strongly localized to the dielectric-semiconductor interface, the electrostatic potential energy can be well approximated by $V(x)=e{\cal E}x$ for $x>0$ with an infinite potential barrier at $x=0$. In this case, the wave function can be written by Airy function with normalization constant such as $\Psi_n(x)=N_nAi({\tilde x}_n)$, where the dimensionless coordinate ${\tilde x}_n=x/a+s_n$ with $s_n$ being the $n$-th zero of Airy function and $a=\left[\hbar^2/\left(2m^*_xe{\cal E}\right)\right]^{1/3}$\cite{MOSFET}. The energy eigenvalues are given by $E_n=-\left(e{\cal E}a\right)s_n$.
The electron number density $n(x)$ can be written by
\begin{equation}
n(x)=\left(\frac{m^*_xk_BT}{\pi\hbar^2}\right)\sum_{n=1}^\infty\ln\left(1+e^{\left(E_{\rm F}-E_n\right)/k_B T}\right)N_n^2 Ai^2\left(x/a+s_n\right).
\end{equation}
From Eq. \ref{Poisson}, one can show that the electric field at near $x=0$ is given by the following formula
\begin{equation}
{\cal E}={e\over \epsilon_s}\int_0^{W_d}dx^\prime\left[n(x^\prime)+N_a\right],
\label{EField}
\end{equation}
where the width of the depletion region becomes maximum such that $W_d\cong\left(2\epsilon_sE_g/e^2N_a\right)^{1/2}$ in the strong inversion regime.
We have solved the Schr{\"o}dinger-Poisson equation, which is a coupled equations of Eq. \ref{Schroedinger} and Eq. \ref{Poisson}.
From now on we will choose the active layer as a p-type GaAs with the following set of parameters: $\epsilon_s=12.9$, $N_a=1.0\times10^{15}{\rm cm}^{-3}$, $\nu=3.5\times10^{12}{\rm Hz}$, $m_x^{*}=m_c^{*}=0.063m_e$, and $W_d=1.4{\rm\mu m}$ taken for strong inversion.
We have obtained ${\cal E}=0.36{\rm MeV/cm}$ for $E_{\rm F}=0.3{\rm eV}$ and ${\cal E}=0.84{\rm MeV/cm}$ for $E_F=0.6{\rm eV}$. In Fig. \ref{Figure1_2}(a), the solid (dashed) curve describes the electron number density $n(x)$ as a function of $-x$ for $E_{\rm F}=0.3{\rm eV} (E_{\rm F}=0.6{\rm eV})$.
In Fig. \ref{Figure1_2}(b), the real part of dielectric permittivity $Re\left[\epsilon(x)\right]$ is plotted as a function of $-x$ in the solid (dashed) curve for $E_{\rm F}=0.3{\rm eV} (E_{\rm F}=0.6{\rm eV})$. One can notice that for $E_{\rm F}=0.6{\rm eV}$, $Re\left[\epsilon(x)\right]$ becomes negative for certain regions of $x$.

\begin{figure}
\epsfxsize=4in \epsfysize=3.5in \epsffile{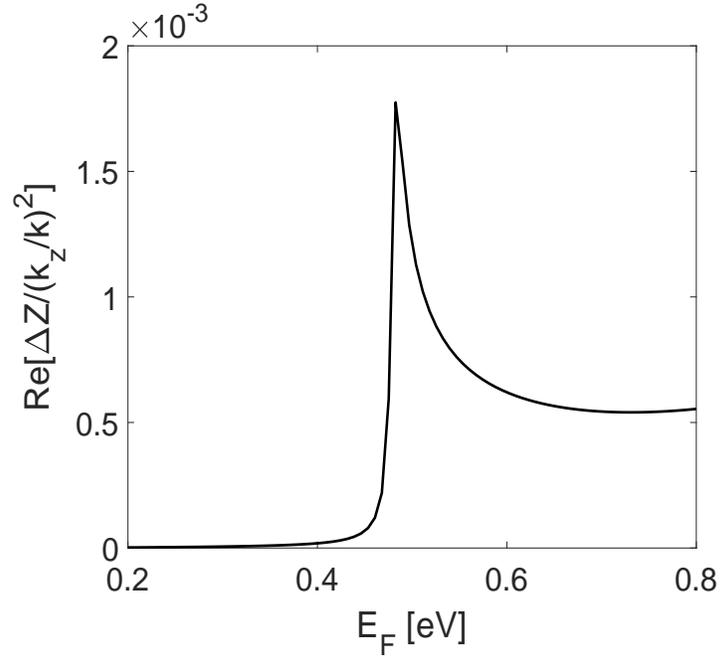}
\caption{
The real part of impedance change at the top of the p-type GaAs as a function of the Fermi energy $E_{\rm F}$ for $\lambda=8\mu m$. It demonstrates a dramatic increase of $Re{\left[\Delta Z/\left({k_z/k}\right)^2\right]}$ above $E_{\rm F}^*\cong 0.48{\rm eV}$.}
\label{Figure1_3}
\end{figure}

At the critical value of $E_{\rm F}^*$, a transition occurs locally from dielectric to metal, where the real part of local dielectric permittivity changes sign.
When $E_{\rm F}$ increases above a critical value of $E_{\rm F}^*$, $Re\left[\epsilon(x)\right]$ crosses to zero at two positions $x=x_1, x_2$ in the inversion layer, which will induce a local plasma resonance. This will significantly change the impedance of the active layer, since the the dominant contribution will come from the term  $-ik^2_{z}/\left[k\epsilon(\omega,x)\right]$ in Eq. \ref{impedance}\cite{akimov}.
At the top of the active layer, the real part of the impedance change $\Delta Z=Z(0)-Z_0$ from the bulk value $Z_0$ can be obtained as follows
\begin{equation}
Re{\left[\Delta Z/\left({k_z/k}\right)^2\right]}=Re{\left[-ik\int^{0}_{-L} dx {1\over \epsilon(\omega,x)}\right]}={\pi k\over \epsilon_\infty}\sum_{i=1,2}\left|\frac{d\ln n}{dx_i}\right|^{-1}
\end{equation}
where $Z_0=(i/\epsilon_\infty)\sqrt{k^2_{z}/k^2-\epsilon_\infty}$ and we have assumed that $\omega\gg\nu$ and $L>W_d$.
We have numerically calculated $Re{\left[\Delta Z/\left({k_z/k}\right)^2\right]}$ as a function of the Fermi energy $E_{\rm F}$ for $\lambda=8\mu m$. Fig. \ref{Figure1_3} demonstrates a dramatic increase of $Re{\left[\Delta Z/\left({k_z/k}\right)^2\right]}$ above $E_{\rm F}^*\cong 0.48{\rm eV}$.

Now we want to show that the transmission of the SPP will be suppressed due to the coupling to the local plasma resonance above $E_{\rm F}^*$.
The impedance $Z(x)$ is a continuous function of $x$ except for the graphene layer. The boundary conditions for $H_y(x)$ and $E_z(x)$ at the graphene layer will yield the following boundary condition for the impedance
\begin{equation}
Z^{-1}(d^+)-Z^{-1}(d^-)=\frac{\sigma_{\Sigma}(\omega)}{\epsilon_{0}c}
\end{equation}
where $d$ is the thickness of the h-BN layer, $Z(d^+)$ represents the impedance right above the graphene layer, and $Z(d^-)$ the impedance right below the graphene layer. Here $Z(d^+)$ will be described by the impedance of air, which can be written by $Z(d^+)=-(i/\epsilon_{0})\sqrt{k^2_{z}/k^2-\epsilon_{0}}$. In the terahertz and mid-infrared regime, the optical conductivity of graphene $\sigma_{\Sigma}(\omega)$ can be well described by the following formula
\begin{equation}
\sigma_{\Sigma}(\omega)=\frac{e^2\epsilon_{F}}{\pi\hbar^2}\frac{i}{\omega+i\tau^{-1}}
\end{equation}
where the Fermi energy of the graphene is given by $\epsilon_{\rm F}=\hbar v_{\rm F}\sqrt{\pi n}$ with $n$ and $v_{\rm F}$ being the electron number density and the Fermi velocity, respectively. For the exfoliated graphene, it has been experimentally demonstrated that $v_{\rm F}\cong10^6{\rm m/s}, \mu_e\cong10^4{\rm cm}^2/Vs$, and $\tau=\mu_e\epsilon_{\rm F}/\left(ev_{\rm F}^2\right)$\cite{akgeim}.

When the h-BN layer is thick enough to neglect the coupling to the active layer, one can approximate that $Z(d^-)\cong(i/\epsilon_{\rm BN})\sqrt{k^2_{z}/k^2-\epsilon_{\rm BN}}$.
In this limit, the dispersion relation of the SPP can be obtained by solving the following equation
\begin{equation}
\frac{\epsilon_{0}}{\sqrt{k_z^2-\epsilon_{0}k^2}}+\frac{\epsilon_{\rm BN}}{\sqrt{k_z^2-\epsilon_{\rm BN}k^2}}= \frac{4\alpha\epsilon_{\rm F}}{\hbar k}\frac{1}{\omega+i\tau^{-1}}
\label{SurfacePlasmon}
\end{equation}
which agrees exactly with the equation obtained by Jablan {\em et al.}\cite{jablan}.
For the fixed frequency $\omega$ of an incident light, one can obtain the complex-valued wave number $k_z$ by solving Eq. \ref{SurfacePlasmon}.
The SPPs thus obtained are a type of surface waves, guided along the interface in much the same way as light can be guided by an optical fiber. Their wavelengths
are much shorter than that of an incident light. Hence, they can have tighter spatial confinement and higher local field intensity.
Perpendicular to the interface, they have sub-wavelength-scale confinement. The SPP will propagate along the interface until its energy is lost either
by absorption in the metal or scattering into other directions.

The wavelength of the SPP is given by $\lambda_{\rm P}=2\pi/Re\left[k_z\right]$.
The propagation length of the SPP is defined as the distance for the SPP intensity to decay by a factor of $1/e$, which can be written by $l_{\rm P}=1/\left(2Im \left[k_z\right]\right)$. Since $\vert k_z/k\vert\gg 1$, the dispersion relation and the propagation length of the SPP will be given by $\omega=\left[e^2 \epsilon_{\rm F}/((\epsilon_{0}+\epsilon_{\rm BN})\pi\hbar^2)\right]^{1/2}\sqrt{Re\left[k_z\right]}$ and $l_{\rm P}=\left(\omega\tau/4\pi\right)\lambda_{\rm P}$, respectively.

\begin{figure}
\epsfxsize=6.5in \epsfysize=3.8in \epsffile{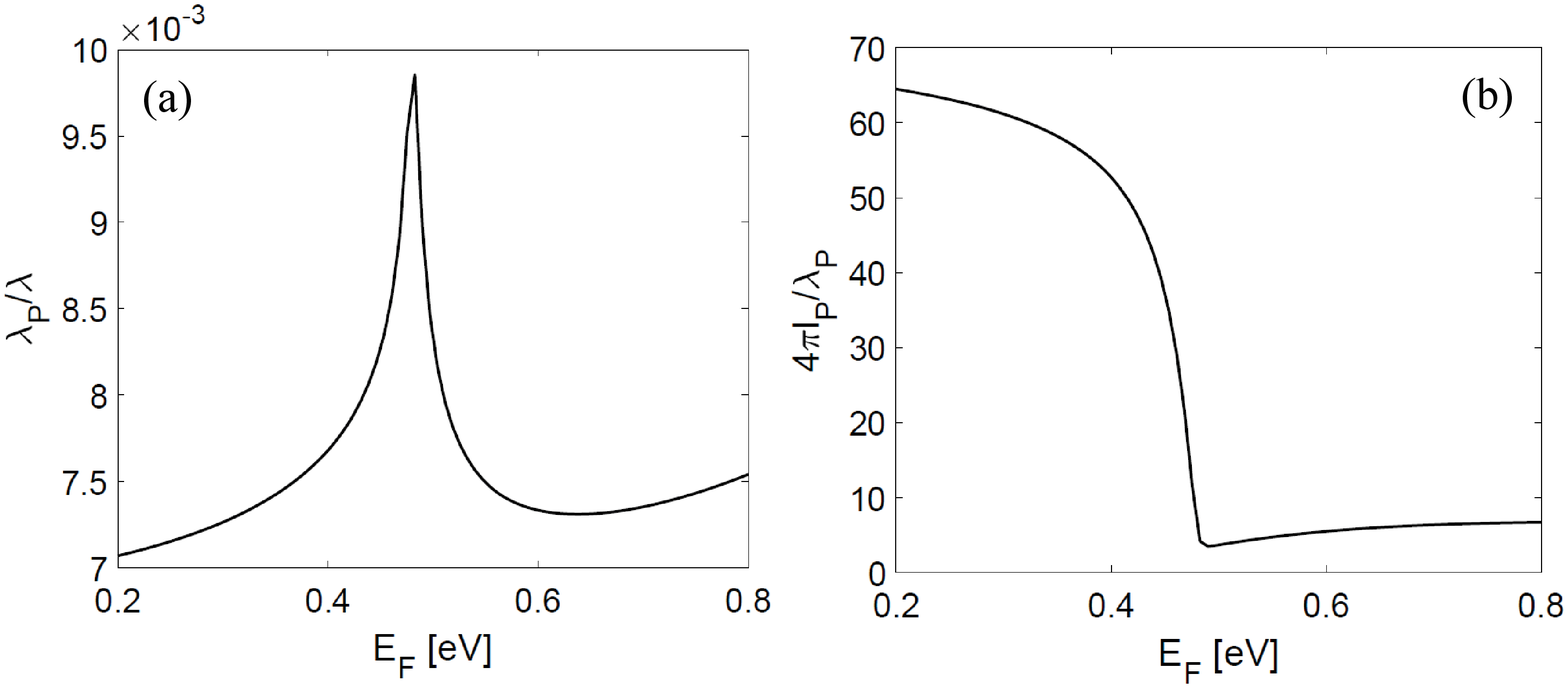}
\caption{(a) The wavelength of the SPP is plotted as a function of $E_{\rm F}$.
For $E_{\rm F}=0.2{\rm eV}$, one can estimate that $\lambda_{\rm P}/\lambda\cong7.0\times10^{-3}$. Hence the wavelength of the SPP can reduce to $\sim1/143$ of that of an incident light.
As $E_{\rm F}$ increases, it will increase, but remain to be on the order of $\sim1/100$ of that of an incident light.
(b) The ratio of the propagation length to the wavelength of the SPP is plotted as a function of $E_{\rm F}$.
One can notice that $4\pi l_{\rm P}/\lambda_{\rm P}$ can be significantly reduced by a factor of $\sim18$ upon switching at $E_{\rm F}^*\cong 0.48{\rm eV}$ compared to the value at $E_{\rm F}=0.2{\rm eV}$.}
\label{Figure2}
\end{figure}

\begin{figure}
\epsfxsize=4.5in \epsfysize=4in \epsffile{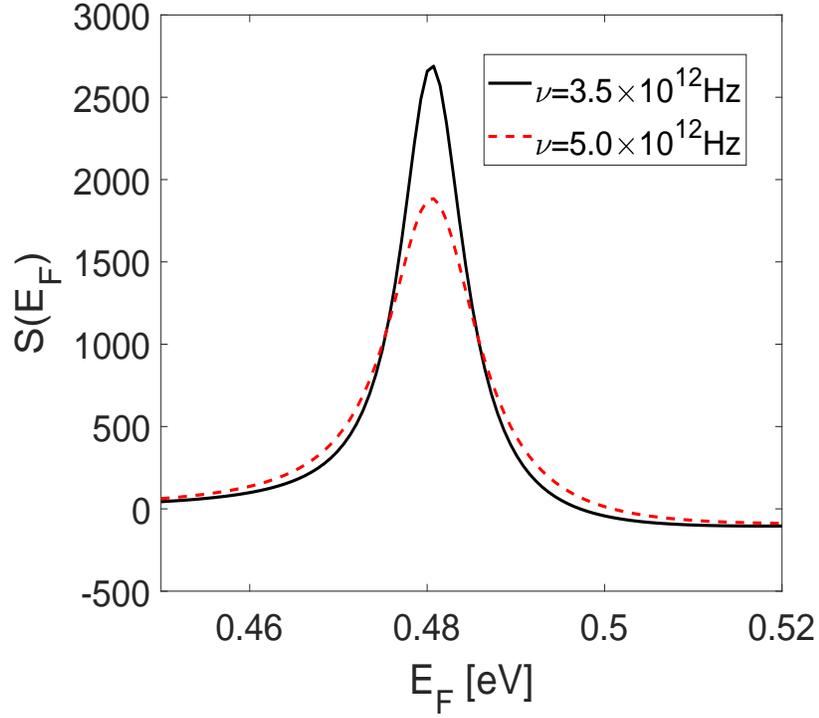}
\caption{
$S(E_{\rm F})$ is plotted as a function of $E_{\rm F}$ for two values of plasma damping rate $\nu=3.5\times10^{12}{\rm Hz}, 5\times 10^{12}{\rm Hz}$.
The sharpness of switching can be measured by Q-factor, which is given by $Q=E_{\rm Fm}/\Delta E_{\rm F}$.   Here $S(E_{\rm F})$ has a maximum value at $E_{\rm Fm}$ and $\Delta E_{\rm F}$ represents the FWHM.
The black solid line describes $S(E_{\rm F})$ for $\nu=3.5\times10^{12}{\rm Hz}$ and $Q\cong 58$. The red dashed line describes $S(E_{\rm F})$ for $\nu=5\times 10^{12}{\rm Hz}$ and $Q\cong 42$.}
\label{Figure3}
\end{figure}

\begin{figure}
\epsfxsize=4.5in \epsfysize=4in \epsffile{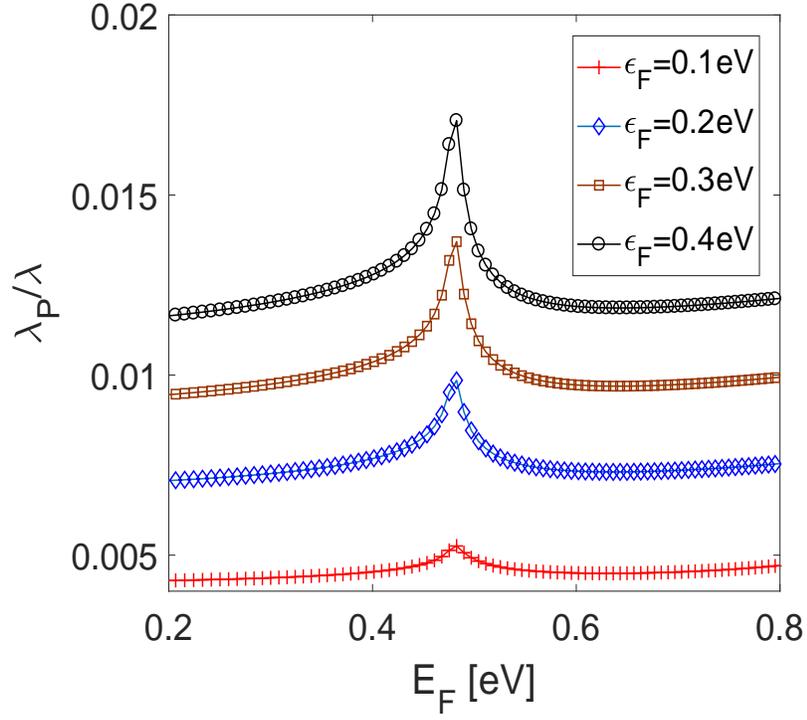}
\caption{
The wavelength of the SPP is plotted as a function of $E_{\rm F}$ for four different values of Fermi energy $\epsilon_{\rm F}$ of graphene. One can notice that $\lambda_{\rm P}$ increases with the increase of $\epsilon_{\rm F}$ at the fixed $E_{\rm F}$.}
\label{Figure4}
\end{figure}

\begin{figure}
\epsfxsize=4.5in \epsfysize=4in \epsffile{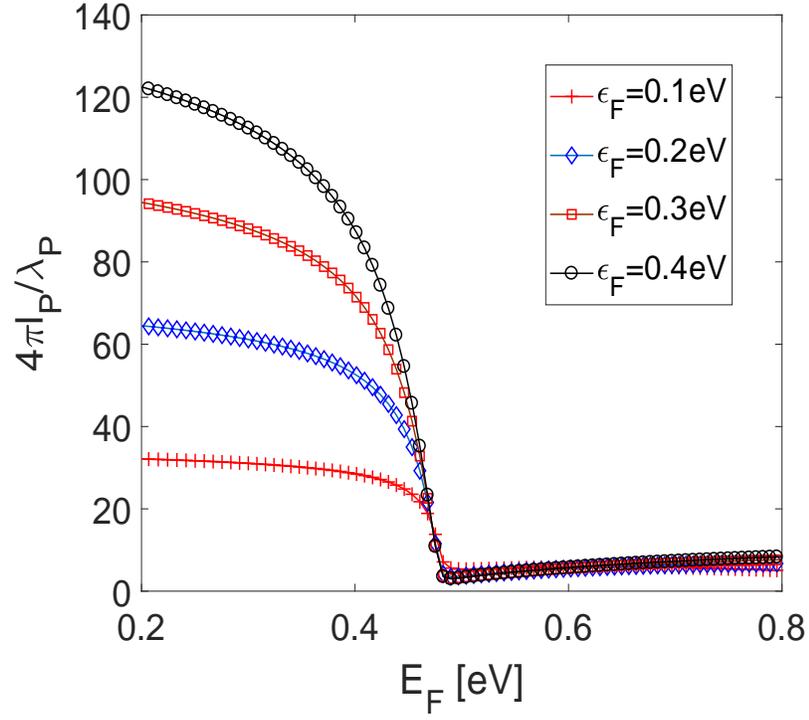}
\caption{
The ratio of the propagation length to the wavelength of the SPP is plotted as a function of $E_{\rm F}$ for four different values of Fermi energy $\epsilon_{\rm F}$ of graphene. One can clearly see that $4\pi l_{\rm P}/\lambda_{\rm P}$ increases with the Fermi energy of graphene below $E_{\rm F}^*$.}
\label{Figure5}
\end{figure}

\begin{figure}
\epsfxsize=7in \epsfysize=4in \epsffile{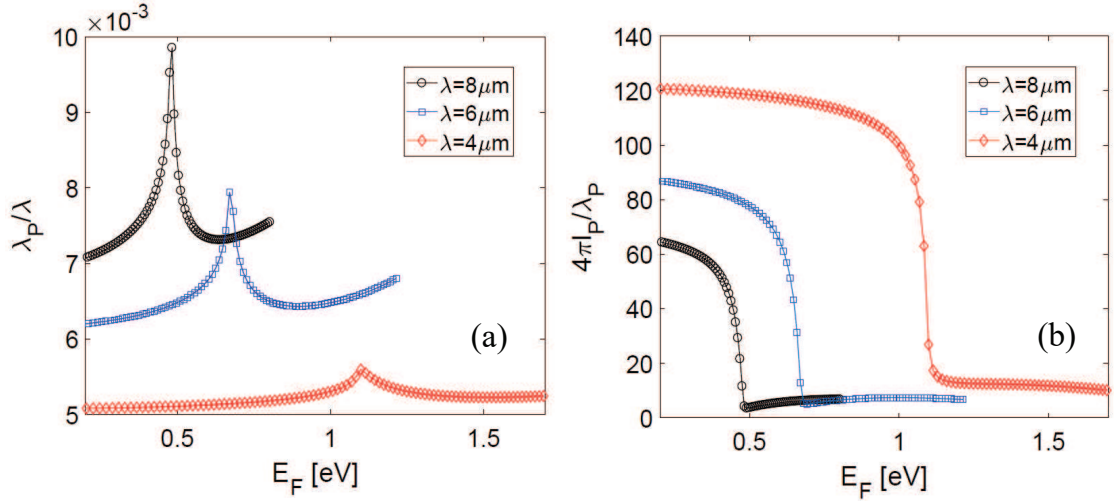}
\caption{
(a) The wavelength of the SPP is plotted as a function of $E_{\rm F}$ for three different values of $\lambda=4,6,8\mu m$.
(b) The ratio of the propagation length to the wavelength of the SPP is plotted as a function of $E_{\rm F}$ for three different values of $\lambda=4,6,8\mu m$.}
\label{Figure6}
\end{figure}

For relatively thin film of the h-BN layer with thickness $d\ll\lambda_{\rm P}$, the SPP will be strongly coupled to the active layer.
For an incident light with mid-infrared wavelength $\lambda=8\mu m$ and the exfoliated graphene with $\epsilon_{\rm F}=0.2{\rm eV}$, we have calculated the complex-valued wave number $k_z$ as a function of the Fermi energy $E_{\rm F}$ by numerically solving Eq. \ref{impedance}. The thickness of the h-BN layer has been taken to be $3{\rm nm}$, which is thick enough to function as a gate dielectric\cite{hBN}, and that of the p-type GaAs layer $L>W_d$.
In Fig. \ref{Figure2}(a), the wavelength of the SPP is plotted as a function of $E_{\rm F}$.
For $E_{\rm F}=0.2{\rm eV}$, one can estimate that $\lambda_{\rm P}/\lambda\cong7.0\times10^{-3}$. Hence the wavelength of the SPP has been reduced to $\sim1/143$ of that of an incident light.
As $E_{\rm F}$ increases, it will increase, but remain to be on the order of $\sim1/100$ of that of an incident light. This makes graphene as a promising material for practical application to sub-wavelength plasmonic devices.
In Fig. \ref{Figure2}(b), the ratio of the propagation length to the wavelength of the SPP is plotted as a function of $E_{\rm F}$.
One can clearly notice that $4\pi l_{\rm P}/\lambda_{\rm P}$ can be significantly reduced by a factor of $\sim18$ upon switching at $E_{\rm F}^*\cong 0.48{\rm eV}$ compared to the value at $E_{\rm F}=0.2{\rm eV}$.
We have also confirmed that the propagation length of the SPP can be reduced by a factor of $\sim15$ upon switching.
This demonstrates that our device can effectively operate as a sub-wavelength plasmonic switch for the SPP.
In order to estimate the sharpness of switching, we will define the function $S(E_{\rm F})$ as follows: $S(E_{\rm F})=\partial Im\left[k_z/k\right]/\partial E_{\rm F}$.
In Fig. \ref{Figure3}, $S(E_{\rm F})$ is plotted as a function of $E_{\rm F}$ for two different values of plasma damping rate $\nu=3.5\times10^{12}{\rm Hz}, 5\times 10^{12}{\rm Hz}$.
The sharpness of switching can be measured by Q-factor, which will be defined by $Q=E_{\rm Fm}/\Delta E_{\rm F}$. Here $S(E_F)$ has a maximum value at $E_{\rm Fm}$ and $\Delta E_{\rm F}$ represents the FWHM.
The black solid line describes $S(E_{\rm F})$ for $\nu=3.5\times10^{12}{\rm Hz}$. One can estimate that $Q\cong 58$. The red dashed line describes $S(E_{\rm F})$ for $\nu=5\times 10^{12}{\rm Hz}$. One can estimate that $Q\cong 42$. We have confirmed that the sharpness of switching line shape mainly depends on the plasma damping rate $\nu$ of the active layer. For the fixed value of $\epsilon_{\rm F}$, the sharpness of switching line shape exhibits only a weak dependence on the electron mobility of graphene. In Fig. \ref{Figure4}, the wavelength of the SPP is plotted as a function of $E_{\rm F}$ for four different values of Fermi energy $\epsilon_{\rm F}$ of graphene. One can notice that $\lambda_{\rm P}$ increases with the increase of $\epsilon_{\rm F}$ at the fixed $E_{\rm F}$. For $E_{\rm F}<E_{\rm F}^*$, this can be easily understood by the following formula $\omega\cong\left[e^2 \epsilon_{\rm F}/((\epsilon_{0}+\epsilon_{\rm BN})\pi\hbar^2)\right]^{1/2}\sqrt{Re\left[k_z\right]}\propto\left(\epsilon_{\rm F}/\lambda_{\rm P}\right)^{1/2}$. Hence $\lambda_{\rm P}$ increases linearly with $\epsilon_{\rm F}$ for the fixed frequency $\omega$ as shown in Fig. \ref{Figure4}. In Fig. \ref{Figure5}, the ratio of the propagation length to the wavelength of the SPP is plotted as a function of $E_{\rm F}$ for four different values of Fermi energy $\epsilon_{\rm F}$ of graphene. One can clearly see that $4\pi l_{\rm P}/\lambda_{\rm P}$ increases with the Fermi energy of graphene below $E_{\rm F}^*$.
Below $E_{\rm F}^*$, this can be easily understood through the following relation $4\pi l_{\rm P}/\lambda_{\rm P}\cong\omega\tau\propto\epsilon_{\rm F}$ since $\tau=\mu_e\epsilon_{\rm F}/\left(ev_{\rm F}^2\right)$. Hence $l_{\rm P}/\lambda_{\rm P}$ increases linearly with $\epsilon_{\rm F}$.
In Fig. \ref{Figure6}(a), the wavelength of the SPP is plotted as a function of $E_{\rm F}$ for three different values of $\lambda=4,6,8\mu m$. One can notice that while the critical value of Fermi energy increases with the decrease of $\lambda$, $\lambda_{\rm P}/\lambda$ decreases with the decrease of $\lambda$.
In Fig. \ref{Figure6}(b), the ratio of the propagation length to the wavelength of the SPP is plotted as a function of $E_{\rm F}$ for three different values of $\lambda=4,6,8\mu m$. One can clearly notice that $4\pi l_{\rm P}/\lambda_{\rm P}$ increases with the decrease of $\lambda$ below $E_{\rm F}^*$.

To summarize, we have proposed a sub-wavelength plasmonic waveguide based on graphene, which can effectively gate the transmission of the SPP by varying gate voltage.
As a van der Waals material, thin h-BN film has been chosen to be a gate dielectric material to protect the high mobility of graphene. In order to achieve
a fast switching of our device, a p-type GaAs has been taken to be an active layer, which has previously demonstrated ultra-high mobility as shown in delta-doped MOSFET structure.
Graphene plasmonics have practical application ranges from terahertz to mid-infrared regime. For an incident light with mid-infrared wavelength $\lambda=8\mu m$, we have demonstrated that the wavelength of the SPP can be reduced below $1/100$ of that of an incident light. By applying a gate voltage, the propagation length of the SPP can be reduced by a factor of $\sim15$ upon switching.
Our theoretical results strongly support that our plasmonic waveguide can effectively operate as a sub-wavelength plasmonic switch for the SPP.

\begin{acknowledgments}
K.~M. wishes to acknowledge Y.~Akimov, S.~Lee, and J.~H.~Kim for useful discussions.
K.~M. wishes to acknowledge the financial support by Basic Science Research Program through the National Research Foundation of Korea(NRF) funded by the Ministry of Education, Science and Technology(NRF-2016R1D1A1B01013756).

\end{acknowledgments}

\newpage

\clearpage
\newpage

\clearpage
\newpage

\clearpage
\newpage

\clearpage

\clearpage

\clearpage

\clearpage

\clearpage
\newpage

\clearpage

\end{document}